\documentclass[sigconf]{acmart}

\usepackage{adjustbox}
\usepackage{multirow}
\usepackage{hyperref}
\usepackage{enumitem}

\AtBeginDocument{%
  }

\setcopyright{acmlicensed}

\copyrightyear{2025}
\acmYear{2025}
\setcopyright{acmlicensed}\acmConference[WWW Companion '25]{Companion Proceedings of the ACM Web Conference 2025}{April 28-May 2, 2025}{Sydney, NSW, Australia}
\acmBooktitle{Companion Proceedings of the ACM Web Conference 2025 (WWW Companion '25), April 28-May 2, 2025, Sydney, NSW, Australia}
\acmDOI{10.1145/3701716.3715498}
\acmISBN{979-8-4007-1331-6/25/04}

\begin{document}

\title{Graph-Sequential Alignment and Uniformity: Toward Enhanced Recommendation Systems}



\author{Yuwei Cao}
\affiliation{%
  \institution{University of Illinois Chicago}
  \city{Chicago}
  \state{IL}
  \country{USA}}
\email{ycao43@uic.edu}

\author{Liangwei Yang}
\affiliation{%
  \institution{Salesforce AI Research}
  \city{Palo Alto}
  \state{CA}
  \country{USA}}
\email{liangwei.yang@salesforce.com}

\author{Zhiwei Liu}
\affiliation{%
  \institution{Salesforce AI Research}
  \city{Palo Alto}
  \state{CA}
  \country{USA}}
\email{zhiweiliu@salesforce.com} 

\author{Yuqing Liu}
\affiliation{%
  \institution{University of Illinois Chicago}
  \city{Chicago}
  \state{IL}
  \country{USA}}
\email{yliu363@uic.edu}

\author{Chen Wang}
\affiliation{%
  \institution{University of Illinois Chicago}
  \city{Chicago}
  \state{IL}
  \country{USA}}
\email{cwang266@uic.edu}

\author{Yueqing Liang}
\affiliation{%
  \institution{Illinois Institute of Technology}
  \city{Chicago}
  \state{IL}
  \country{USA}}
\email{yliang40@hawk.iit.edu}

\author{Hao Peng}
\orcid{0000-0003-0458-5977}
\authornote{Corresponding author}
\affiliation{%
  \institution{Beihang University}
  \city{Beijing}
  \country{China}}
\affiliation{%
  \institution{Hangzhou Innovation Institute of BUAA}
  \city{Hangzhou}
  \state{Zhejiang}
  \country{China}}
\email{penghao@buaa.edu.cn}

\author{Philip S. Yu}
\affiliation{%
  \institution{University of Illinois Chicago}
  \city{Chicago}
  \state{IL}
  \country{USA}}
\email{psyu@uic.edu}
\renewcommand{\shortauthors}{Yuwei Cao et al.}

\begin{abstract}

  Graph-based and sequential methods are two popular recommendation paradigms, each excelling in its domain but lacking the ability to leverage signals from the other. To address this, we propose a novel method that integrates both approaches for enhanced performance. Our framework uses Graph Neural Network (GNN)-based and sequential recommenders as separate submodules while sharing a unified embedding space optimized jointly. To enable positive knowledge transfer, we design a loss function that enforces alignment and uniformity both within and across submodules. Experiments on three real-world datasets demonstrate that the proposed method significantly outperforms using either approach alone and achieves state-of-the-art results. Our implementations are publicly available at \href{https://github.com/YuweiCao-UIC/GSAU.git}{https://github.com/YuweiCao-UIC/GSAU.git}.
\end{abstract}




\maketitle

\section{Introduction}
In recent years, deep learning-based approaches have significantly advanced recommender systems by enabling expressive representations of users and items. Among these, Graph Neural Network (GNN)~\cite{kipf2016semi, hamilton2017inductive}-based recommenders~\cite{wang2019neural, he2020lightgcn} and sequential recommendation systems~\cite{sun2019bert4rec, kang2018self} have emerged as two widely adopted paradigms, each demonstrating state-of-the-art performance. Despite their effectiveness, both methods exhibit inherent limitations: GNN-based recommenders excel at capturing higher-order collaborative filtering signals through convolutions on bipartite user-item graphs but fail to incorporate sequential information. On the other hand, sequential recommenders adeptly model personalized interaction histories but struggle to account for higher-order correlations between users.

Recognizing the respective strengths of GNN-based and sequential recommenders in modeling higher-order and sequential data, we propose integrating both approaches for enhanced performance. Our method is straightforward, utilizing each architecture as separate submodules while maintaining a shared embedding space optimized by both. The main challenge is ensuring effective knowledge transfer between the two submodules to achieve synergy. 
Specifically, GNN-based recommenders often utilize loss functions like Bayesian personalized ranking (BPR)~\cite{rendle2012bpr} for pairwise ranking or InfoNCE~\cite{gutmann2010noise} for contrastive representation learning, while sequential recommenders typically employ cross-entropy (CE) loss~\cite{sun2019bert4rec} or BPR loss for next-item prediction tasks. 
Studies~\cite{cao2023multi} have shown that directly applying these original loss functions during the co-training of different architectures, even with normalization terms, can lead to unstable training processes and erratic gradients due to conflicts in optimization dynamics. 
To address this issue, we design a novel loss function that enforces alignment and uniformity properties~\cite{wang2020understanding} both within and across the submodules. 
The proposed loss enables the GNN-based and sequential encoders to supplement and benefit one another, achieving significant performance improvements. Our contributions are as follows:
\begin{itemize}[leftmargin=*]
    \item We propose a simple yet effective method that co-trains GNN-based and sequential recommenders to enhance performance.
    \item Our custom loss function enforces alignment and uniformity, facilitating effective knowledge transfer between the GNN-based and sequential encoders.
    \item We demonstrate that integrating GNN-based and sequential encoders delivers significantly better performance compared to using either approach independently. Experimental results on three real-world datasets reveal that our proposed method consistently outperforms state-of-the-art graph-based and sequential recommenders by substantial margins. 
\end{itemize}

\section{Related Work}
\noindent \textbf{Integrating GNN-Based and Sequential Recommenders} GNN-based~\cite{wang2019neural, he2020lightgcn} and sequential recommenders~\cite{sun2019bert4rec, kang2018self} excel at modeling bipartite user-item graphs and user-specific interaction sequences, respectively, but often overlook the complementary value of other data. Few works address this limitation \cite{baikalov2024end}: MCLSR~\cite{wang2022multi} enriches sequential recommenders by graph construction, while MRGSRec~\cite{baikalov2024end} combines graph and sequential embeddings via a fusion layer. In contrast, our method employs a simpler design, enabling GNN-based and sequential encoders to share embedding layers without requiring fusion modules. To ensure effective knowledge transfer, we introduce a loss function that enforces alignment and uniformity~\cite{wang2020understanding} within and across encoders, allowing them to augment and enhance each other’s performance.

\noindent \textbf{Alignment and Uniformity in Recommendation} Recent advances in unsupervised contrastive representation learning~\cite{wang2020understanding} highlight two crucial properties for high-quality representations: alignment, ensuring positive pairs are close, and uniformity, distributing instances evenly across a hypersphere. Contrastive losses that optimize these properties have been applied in collaborative filtering (CF)~\cite{wang2022towards,park2023toward}, GNN-based~\cite{yang2023graph, 10.1145/3627673.3679535, DBLP:conf/wsdm/YangLYLWPY24}, and sequential recommendation systems~\cite{chung2023leveraging}, yielding improved performance over traditional losses like BPR~\cite{rendle2012bpr}, InfoNCE~\cite{gutmann2010noise}, and CE~\cite{sun2019bert4rec}. 
Unlike prior studies, this work investigates the use of these properties for the joint optimization of GNN-based and sequential encoders.

\begin{table*}[t]
  \centering
  \begin{adjustbox}{width=1\linewidth}
  \small
    \begin{tabular}{c|cccc|cccc|cccc}
    \hline
    \multirow{2}{*}{Method} & \multicolumn{4}{c}{Beauty} & \multicolumn{4}{c}{Sports} & \multicolumn{4}{c}{Toys} \\
    \cline{2-13}
     & R@10 & R@50 & N@10 & N@50 & R@10 & R@50 & N@10 & N@50 & R@10 & R@50 & N@10 & N@50 \\
    \hline
    LightGCN & 0.0435 & 0.1157 & 0.0216 & 0.0372 & 0.0299 & 0.0818 & 0.0153 & 0.0265 & 0.0425 & 0.1016 & 0.0226 & 0.0355 \\
DirectAU & 0.0492 & 0.1270 & 0.0252 & 0.0421 & 0.0318 & 0.0892 & 0.0165 & 0.0289 & 0.0501 & 0.1160 & 0.0264 & 0.0406 \\
GraphAU & 0.0489 & 0.1300 & 0.0245 & 0.0421 & 0.0323 & 0.0929 & 0.0168 & 0.0299 & 0.0472 & 0.1154 & 0.0248 & 0.0396 \\
BERT4Rec & 0.0390 & 0.1072 & 0.0189 & 0.0335 & 0.0174 & 0.0553 & 0.0086 & 0.0167 & 0.0299 & 0.0813 & 0.0157 & 0.0267 \\
BERT4Rec (u) & 0.0487 & 0.1207 & 0.0258 & 0.0414 & 0.0195 & 0.0604 & 0.0095 & 0.0182 & 0.0471 & 0.1090 & 0.0255 & 0.0387 \\
SASRec & 0.0826 & 0.1752 & 0.0414 & 0.0616 & 0.0467 & 0.1115 & 0.0218 & 0.0359 & \underline{0.0891} & 0.1767 & 0.0434 & 0.0625 \\
SASRec (u) & \underline{0.0833} & \underline{0.1806} & \underline{0.0443} & \underline{0.0653} & 0.0470 & 0.1137 & 0.0239 & 0.0383 & 0.0878 & \underline{0.1799} & \underline{0.0453} & \underline{0.0653} \\
MRGSRec \textsuperscript{1} & 0.0681 & / & 0.0349 & / & 0.0434 & / & 0.0227 & / & / & / & / & / \\
\hline
GSAU & 0.0631 & 0.1622 & 0.0303 & 0.0518 & \underline{0.0491} & \underline{0.1188} & \underline{0.0256} & \underline{0.0406} & 0.0655 & 0.1653 & 0.0306 & 0.0524 \\
GSAU (rec) & \textbf{0.0876} & \textbf{0.1974} & \textbf{0.0453} & \textbf{0.0694} & \textbf{0.0533} & \textbf{0.1285} & \textbf{0.0272} & \textbf{0.0434} & \textbf{0.0937} & \textbf{0.1966} & \textbf{0.0474} & \textbf{0.0698} \\
\hline
$\Delta$(\%) & 5.16 & 9.30 & 2.26 & 6.28 & 13.40 & 13.02 & 13.81 & 13.32 & 5.16 & 9.28 & 4.64 & 6.89 \\
    \hline
  \end{tabular}
  \end{adjustbox}
  \caption{Overall performance. The best (second-best) results are in bold (underlined). \textsuperscript{1} marks results from~\cite{baikalov2024end}.}
  \label{table:overall}
\end{table*}

\begin{table}[t]
  \centering
  \begin{adjustbox}{width=0.94\linewidth}
  \small
    \begin{tabular}{l|cccc}
    \hline
     Method & R@10 & R@50 & N@10 & N@50 \\
    \hline
GSAU (rec) & \textbf{0.0876} & \textbf{0.1974} & \textbf{0.0453} & \textbf{0.0694} \\
\hline
w/o graph & \underline{0.0833} & \underline{0.1806} & \underline{0.0443} & \underline{0.0653} \\
w/o sequential & 0.0492 & 0.1270 & 0.0252 & 0.0421 \\
w/o rec & 0.0631 & 0.1622 & 0.0303 & 0.0518 \\ 
w/o u-i uniform & 0.0739 & 0.1743 & 0.0391 & 0.0609 \\
replace sequential & 0.0526 & 0.1397 & 0.0256 & 0.0444 \\
    \hline
  \end{tabular}
  \end{adjustbox}
  \caption{Ablation study on Beauty.}
  \label{table:ablation}
\end{table}

\section{Methodology}

\begin{figure}[ht]
\begin{center}
\centerline{\includegraphics[width=0.92\columnwidth]{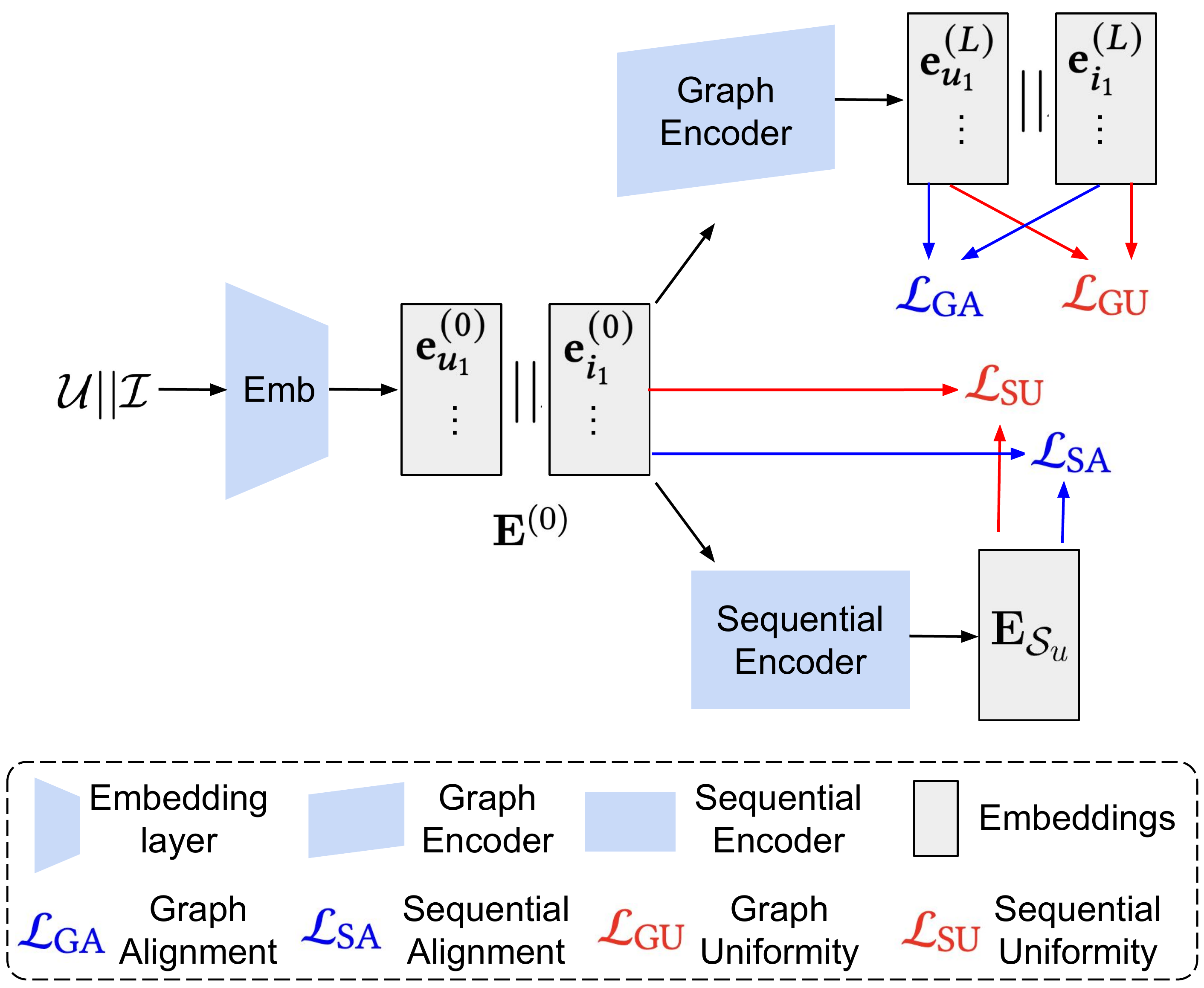}}
\caption{Overview of the proposed GSAU model.}
\label{figure:framework}
\end{center}
\end{figure}

We introduce the architecture of our proposed model and detail its loss function that enforces alignment and uniformity. The method
is named as graph-sequential alignment and uniformity (GSAU). 
An overview of GSAU is provided in Figure~\ref{figure:framework}.
\subsection{Model Architecture}
We adopt a simple approach that retains the original architectures of the GNN-based and sequential encoders, leveraging a shared embedding layer to enable knowledge sharing between them. 

\subsubsection{Shared Embedding Layer}
\label{sec:embed_layer}
Let $\mathcal{U} = \{u_n\}_{n=1}^{N}$ and $\mathcal{I} = \{i_m\}_{m=1}^{M}$ denote the user and item set, respectively. An embedding layer, denoted as $\text{Emb}(\cdot)$, is a network that maps each user and item into a low-dimensional representation $\text{Emb}(u), \text{Emb}(i) \in \mathbb{R}^d$, where $d$ is the dimension of the latent space. 
The embedding layer maps all user and items to their representations, denoted as $\mathbf{E}^{(0)} = (\mathbf{e}_{u_1}^{(0)}, ..., \mathbf{e}_{i_1}^{(0)}, ...) = (\text{Emb}(u_1),..., \text{Emb}(i_1),...) \in \mathbb{R}^{(N+M)\times d}$.
\subsubsection{GNN Encoder}
The core principle of a GNN encoder is to learn node representations through propagating features across the graph~\cite{he2020lightgcn}. This is achieved through iteratively aggregating information from neighboring nodes, as formulated by:
\begin{equation}
\mathbf{e}_{u}^{(l+1)} = \text{GraphAGG}^{(l+1)}\big(\mathbf{e}_{u}^{(l)}, \{\mathbf{e}_{i}^{(l)} \mid (u, i) \in \mathcal{R}\}\big),
\end{equation}
\begin{equation}
\mathbf{e}_{i}^{(l+1)} = \text{GraphAGG}^{(l+1)}\big(\mathbf{e}_{i}^{(l)}, \{\mathbf{e}_{u}^{(l)} \mid (u, i) \in \mathcal{R}\}\big),
\end{equation}
where $\mathcal{R} = \{(u,i)|u \text{ interacted with } i\}$ is a set of observed user-item interactions. $l=1,...,L$ is the layer number. $\mathbf{e}_{u}^{(l)}$ and $\mathbf{e}_{i}^{(l)}$ are the embeddings of user $u$ and item $i$ at the $l$-th layer. $\text{GraphAGG}(\cdot)$, the graph aggregator, is a aggregating function that summarizes the representations of the target node and its neighbors. Notable examples of aggregators include the weighted sum aggregators used in GIN~\cite{xu2018powerful} and LightGCN~\cite{he2020lightgcn}, the LSTM aggregator from GraphSAGE~\cite{hamilton2017inductive}, and the attention-based aggregator from GAT~\cite{velivckovic2017graph}.

\subsubsection{Sequential Encoder}
A sequential recommender identifies patterns in users' historical interactions. Let $\mathcal{S}_u = [i_{1}^{u}, ...,i_{t}^{u},...,i_{|\mathcal{S}_{u}|}^{u}]$ denote the interacted items in chronological order for $u \in \mathcal{U}$, where $i_{t}^{u} \in \mathcal{I}$ is the item that $u$ has interacted with at time step $t$. Let $\textbf{E}_{\mathcal{S}_{u}}^{(0)} \in \mathbb{R}^{|\mathcal{S}_{u}|\times d}$ denote the initial embedding of $\mathcal{S}_u$ given by the embedding layer. The embedding process of a sequential encoder can be outlined as follows:
\begin{equation}
\mathbf{E}_{\mathcal{S}_{u}} = \text{SequentialAGG}\big(\text{Mask}(\mathbf{E}_{\mathcal{S}_{u}}^{(0)}) + \mathbf{P}_{\mathcal{S}_{u}}\big),
\end{equation}
where $\textbf{P}_{\mathcal{S}_{u}}\in \mathbb{R}^{|\mathcal{S}_{u}|\times d}$ is the positional embedding~\cite{sun2019bert4rec} of $\mathcal{S}_{u}$ that enables the usage of the sequential information. $\text{Mask}(\cdot)$ constructs the cloze task~\cite{sun2019bert4rec} in which a proportion of $\mathcal{S}_{u}$, denoted as $\mathcal{S}_{u}^{m}$, are masked out and are to be inferred from the remaining part of $\mathcal{S}_{u}$, i.e., $\mathcal{S}_{u}^{\setminus m}$. In the case of single-direction attention-based recommenders~\cite{kang2018self}, the masked subset typically includes the right-most item. $\text{SequentialAGG}(\cdot)$ denotes the sequential aggregator, examples include the GRU layers used in GRU4Rec~\cite{hidasi2015session}, as well as the Transformer layers used in SASRec~\cite{kang2018self} and BERT4Rec~\cite{sun2019bert4rec}.

\begin{figure}[ht]
\begin{center}
\centerline{\includegraphics[width=\columnwidth]{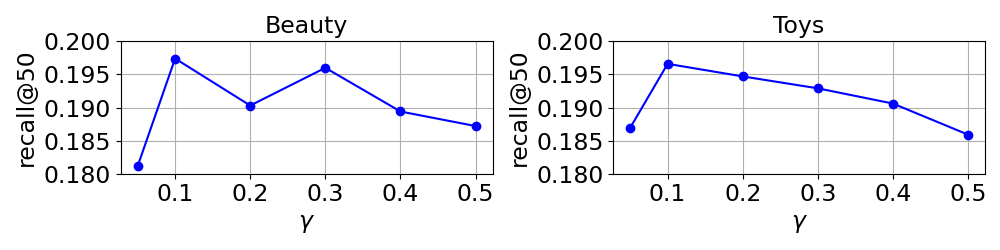}}
\caption{Parameter sensitivity to $\gamma$.}
\label{figure:gamma}
\end{center}
\end{figure}

\subsection{Alignment and Uniformity Loss}
Effective positive knowledge transfer between the GNN and sequential encoders is crucial for obtaining high-quality representations. However, using heterogeneous loss functions (e.g., BPR loss for the GNN encoder and CE loss for the sequential encoder) has been shown~\cite{cao2023multi} to cause instability during model training and degrade representation quality. To mitigate this, we propose loss terms that enforce alignment and uniformity~\cite{wang2020understanding} between the encoders. Given the distribution of the positive pairs $p_{\text{pos}}(\cdot,\cdot)$, the alignment loss follows:
\begin{equation}
\mathcal{L}_{A} \triangleq \underset{(\mathbf{e}_{x},\mathbf{e}_{x^{+}}) \sim p_{\text{pos}}}{\mathbb{E}} \|f(\mathbf{e}_{x}) - f(\mathbf{e}_{x^{+}})\|^{2},
\end{equation}
where $f(\cdot)$ denotes the $l_2$ norm of the input embedding. On the other hand, given the distribution of all instances $p_{\text{data}}$, the uniformity loss is formalized as:
\begin{equation}
\label{eq:uniformity}
\mathcal{L}_{U} \triangleq \log \underset{(\mathbf{e}_{x},\mathbf{e}_{x'}) \overset{\text{i.i.d.}}{\sim} p_{\text{data}}}{\mathbb{E}} \exp\big(-2\|f(\mathbf{e}_{x}) - f(\mathbf{e}_{x'})\|^{2}\big).
\end{equation}

As suggested by their mathematical formulations, the alignment and uniformity losses aim to achieve two objectives: bringing positive instances closer together and dispersing random instances uniformly across a hypersphere.

For the GNN encoder, both items and users are treated as instances. Specifically, the alignment objective encourages users to align closely with the items they have interacted with, while the uniformity objective ensures that the representations of all item and user instances are evenly distributed. Accordingly, we adopt graph alignment loss $\mathcal{L}_{\text{GA}}$ that samples positive pairs from:
\begin{equation}
P_{\text{g\_pos}} = \{(\mathbf{e}_{u}^{(L)}, \mathbf{e}_{i}^{(L)}) \mid (u, i) \in \mathcal{R}\},
\end{equation}
while the random instances for calculating the graph uniformity loss, $\mathcal{L}_{\text{GU}}$, are sampled from:
\begin{equation}
P_{\text{g\_data}} = \{\mathbf{e}_{i}^{(L)}, \mathbf{e}_{i'}^{(L)} \mid i, i' \in \mathcal{I}\} \cup \{\mathbf{e}_{u}^{(L)}, \mathbf{e}_{u'}^{(L)} \mid u, u' \in \mathcal{U}\}.
\end{equation}

For the sequential encoder, we treat items and users’ historical interaction sequences as instances. We adopt sequential alignment loss $\mathcal{L}_{\text{SA}}$ that samples positive pairs from:
\begin{equation}
P_{\text{s\_pos}} = \{(\mathbf{e}_{i}^{(0)}, \mathbf{E}_{\mathcal{S}_{u}}) \mid i \in \mathcal{S}_{u}^{m}\},
\end{equation}
while the random instances for calculating the sequential uniformity loss, $\mathcal{L}_{\text{SU}}$, are sampled from:
\begin{equation}
\begin{split}
P_{\text{s\_data}} = &\ \{\mathbf{e}_{i}^{(0)}, \mathbf{e}_{i'}^{(0)} \mid i, i' \in \mathcal{I}\} \\
&\ \cup \{\mathbf{E}_{\mathcal{S}_{u}}, \mathbf{E}_{\mathcal{S}_{u'}} \mid u, u' \in \mathcal{U}\} \\
&\ \cup \{\mathbf{e}_{i}^{(0)}, \mathbf{E}_{\mathcal{S}_{u}} \mid i \not\in \mathcal{S}_{u}\}.
\end{split}
\label{eq:p_s_data}
\end{equation}

Note that unlike items and users, which have their own dedicated embedding layers (Section~\ref{sec:embed_layer}), the embeddings for users’ historical interaction sequences are derived from item embeddings. Consequently, these interaction sequences act as `hyper-items,' mapped within the item embedding space. To further unify this space, we introduce a third subset in Equation (\ref{eq:p_s_data}). This subset encourages sequences to diverge from negative items, complementing the first two subsets that promote dispersion of items and sequences independently.
The overall loss function is as follows:
\begin{equation}
\mathcal{L} = \mathcal{L}_{\text{GA}} + \mathcal{L}_{\text{SA}} + \gamma \big(\mathcal{L}_{\text{GU}} + \mathcal{L}_{\text{SU}}\big),
\end{equation}
The weight $\gamma$ determines the desired balance between alignment and uniformity. All sampling is conducted at the batch level.



\section{Experiment}

\subsection{Experimental Setup}
\noindent \textbf{Datasets and Evaluation Metrics.} We conduct experiments on three real-world datasets: Amazon Toys \& Games, Beauty, and Sports \& Outdoors. Following~\cite{baikalov2024end, chung2023leveraging, cao2024aligning}, we retain the 5-core data and adopt a leave-one-out evaluation strategy. Specifically, for each user's historical interaction sequence (with interactions sorted in ascending order by timestamp), the last interaction is used for testing, the second-to-last for validation, and the remaining interactions for training. 
We report top-k recall and Normalized
Discounted Cumulative Gain (NDCG), where
k is set to 10/50.

\noindent \textbf{Models and Implementation Details.} We evaluate \textbf{GSAU} and its variant, \textbf{GSAU (rec)}, which replaces the sequential alignment loss, $\mathcal{L}_{\text{SA}}$, with recommendation loss (i.e., CE). Comparisons are made with graph-based models like \textbf{LightGCN}~\cite{he2020lightgcn}, \textbf{DirectAU}~\cite{wang2022towards} (applying alignment and uniformity loss on LightGCN), and \textbf{GraphAU}~\cite{yang2023graph} (enhancing DirectAU with higher-order alignment). We also compare with sequential recommenders, including \textbf{BERT4Rec}~\cite{sun2019bert4rec}, \textbf{SASRec}~\cite{kang2018self}, and their uniformity-enhanced variants~\cite{chung2023leveraging}, \textbf{BERT4Rec (u)} and \textbf{SASRec (u)}. Additionally, we evaluate against \textbf{MRGSRec}~\cite{baikalov2024end}, a graph-sequential method. For GSAU and GSAU~(rec), we use LightGCN as the graph encoder, SASRec as the sequential encoder, and tune $\gamma$ in \{0.05, 0.1, 0.2, 0.3, 0.4, 0.5\}. For a fair comparison, all methods are implemented using the RecBole framework~\cite{zhao2021recbole}. We adopt the Adam optimizer, with a learning rate of 1e-3, embedding size of 64, and a maximum of 300 epochs, applying early stopping if validation NDCG@20 drops for 10 consecutive epochs. 


\subsection{Overall Performance} As shown in Table~\ref{table:overall}, GSAU (rec) outperforms the best baseline, SASRec (u), by notable margins across datasets and metrics ($\Delta$ \%). This demonstrates that integrating graph and sequential models delivers superior performance compared to using either model independently, with GSAU (rec) offering an effective approach to achieve this synergy. Comparing GSAU with GSAU (rec), replacing the sequential alignment loss with CE loss improves performance. This is likely because CE loss not only aligns sequences with their positive items but also disperses them from all negatives, enhancing uniformity of the item embedding space alongside alignment.

\subsection{Ablation and Hyperparameter Sensitivity Analysis}

Table~\ref{table:ablation} shows the ablation study results. The degraded performance of \textit{w/o graph} (removing the graph encoder) and \textit{w/o sequential} (removing the sequential encoder) highlights the importance of both encoders. Similarly, the poorer results of \textit{w/o rec} (using alignment loss instead of CE loss in the sequential encoder) and \textit{w/o u-i uniform} (removing the third subset in Equation~\ref{eq:p_s_data}) demonstrate the importance of dispersing sequences from the negative items and fully unifying the item embedding space. The results of \textit{replace sequential} (using BERT4Rec instead of SASRec as the sequential encoder) confirm the value of a strong sequential encoder and show that the proposed method remains effective regardless of the choice of the sequential encoder, outperforming BERT4Rec and BERT4Rec (u) in Table\ref{table:overall}.
Figure~\ref{figure:gamma} shows that a $\gamma$ value between 0.1 and 0.4 strikes a balance between alignment and uniformity losses, yielding good performance, with $\gamma=0.1$ achieving the best results. We also observe notable performance deterioration (not shown here) when $\gamma$ is too large (> 0.5) or too small (< 0.05).

\section{Conclusion}
We propose a simple and effective method that combines GNN-based and sequential recommenders, leveraging a custom loss function to enforce alignment and uniformity for efficient knowledge transfer. Experiments show significant performance gains over using GNN-based or sequential recommenders alone.

\begin{acks}
This work is supported in part by NSF under grants III-2106758, and POSE-2346158.
Hao Peng is supported by the National Key R\&D Program of China through grant 2022YFB3104703, the NSFC through grants 62322202 and 62441612, and the ``Pioneer'' and ``Leading Goose'' R\&D Program of Zhejiang through grant 2025C02044.
\end{acks}
\bibliographystyle{ACM-Reference-Format}
\bibliography{sample-base}


\begin{thebibliography}{23}


\ifx \showCODEN    \undefined \def \showCODEN     #1{\unskip}     \fi
\ifx \showDOI      \undefined \def \showDOI       #1{#1}\fi
\ifx \showISBNx    \undefined \def \showISBNx     #1{\unskip}     \fi
\ifx \showISBNxiii \undefined \def \showISBNxiii  #1{\unskip}     \fi
\ifx \showISSN     \undefined \def \showISSN      #1{\unskip}     \fi
\ifx \showLCCN     \undefined \def \showLCCN      #1{\unskip}     \fi
\ifx \shownote     \undefined \def \shownote      #1{#1}          \fi
\ifx \showarticletitle \undefined \def \showarticletitle #1{#1}   \fi
\ifx \showURL      \undefined \def \showURL       {\relax}        \fi
\providecommand\bibfield[2]{#2}
\providecommand\bibinfo[2]{#2}
\providecommand\natexlab[1]{#1}
\providecommand\showeprint[2][]{arXiv:#2}

\bibitem[Baikalov and Frolov(2024)]%
        {baikalov2024end}
\bibfield{author}{\bibinfo{person}{Vladimir Baikalov} {and} \bibinfo{person}{Evgeny Frolov}.} \bibinfo{year}{2024}\natexlab{}.
\newblock \showarticletitle{End-to-End Graph-Sequential Representation Learning for Accurate Recommendations}. In \bibinfo{booktitle}{\emph{Companion Proceedings of the ACM on Web Conference 2024}}. \bibinfo{pages}{501--504}.
\newblock


\bibitem[Cao et~al\mbox{.}(2024)]%
        {cao2024aligning}
\bibfield{author}{\bibinfo{person}{Yuwei Cao}, \bibinfo{person}{Nikhil Mehta}, \bibinfo{person}{Xinyang Yi}, \bibinfo{person}{Raghunandan Keshavan}, \bibinfo{person}{Lukasz Heldt}, \bibinfo{person}{Lichan Hong}, \bibinfo{person}{Ed~H Chi}, {and} \bibinfo{person}{Maheswaran Sathiamoorthy}.} \bibinfo{year}{2024}\natexlab{}.
\newblock \showarticletitle{Aligning Large Language Models with Recommendation Knowledge}.
\newblock \bibinfo{journal}{\emph{arXiv preprint arXiv:2404.00245}} (\bibinfo{year}{2024}).
\newblock


\bibitem[Cao et~al\mbox{.}(2023)]%
        {cao2023multi}
\bibfield{author}{\bibinfo{person}{Yuwei Cao}, \bibinfo{person}{Liangwei Yang}, \bibinfo{person}{Chen Wang}, \bibinfo{person}{Zhiwei Liu}, \bibinfo{person}{Hao Peng}, \bibinfo{person}{Chenyu You}, {and} \bibinfo{person}{Philip~S Yu}.} \bibinfo{year}{2023}\natexlab{}.
\newblock \showarticletitle{Multi-task Item-attribute Graph Pre-training for Strict Cold-start Item Recommendation}.
\newblock \bibinfo{journal}{\emph{arXiv preprint arXiv:2306.14462}} (\bibinfo{year}{2023}).
\newblock


\bibitem[Chung et~al\mbox{.}(2023)]%
        {chung2023leveraging}
\bibfield{author}{\bibinfo{person}{Hyunsoo Chung}, \bibinfo{person}{AI Omnious}, {and} \bibinfo{person}{Jungtaek Kim}.} \bibinfo{year}{2023}\natexlab{}.
\newblock \showarticletitle{Leveraging Uniformity of Normalized Embeddings for Sequential Recommendation}.
\newblock \bibinfo{journal}{\emph{History}} \bibinfo{volume}{4}, \bibinfo{number}{5} (\bibinfo{year}{2023}).
\newblock


\bibitem[Gutmann and Hyv{\"a}rinen(2010)]%
        {gutmann2010noise}
\bibfield{author}{\bibinfo{person}{Michael Gutmann} {and} \bibinfo{person}{Aapo Hyv{\"a}rinen}.} \bibinfo{year}{2010}\natexlab{}.
\newblock \showarticletitle{Noise-contrastive estimation: A new estimation principle for unnormalized statistical models}. In \bibinfo{booktitle}{\emph{Proc. AISTATS}}.
\newblock


\bibitem[Hamilton et~al\mbox{.}(2017)]%
        {hamilton2017inductive}
\bibfield{author}{\bibinfo{person}{Will Hamilton}, \bibinfo{person}{Zhitao Ying}, {and} \bibinfo{person}{Jure Leskovec}.} \bibinfo{year}{2017}\natexlab{}.
\newblock \showarticletitle{Inductive representation learning on large graphs}. In \bibinfo{booktitle}{\emph{Proceedings of the NeurIPS}}. \bibinfo{pages}{1024--1034}.
\newblock


\bibitem[He et~al\mbox{.}(2020)]%
        {he2020lightgcn}
\bibfield{author}{\bibinfo{person}{Xiangnan He}, \bibinfo{person}{Kuan Deng}, {et~al\mbox{.}}} \bibinfo{year}{2020}\natexlab{}.
\newblock \showarticletitle{Lightgcn: Simplifying and powering graph convolution network for recommendation}. In \bibinfo{booktitle}{\emph{Proc. SIGIR}}.
\newblock


\bibitem[Hidasi(2015)]%
        {hidasi2015session}
\bibfield{author}{\bibinfo{person}{B Hidasi}.} \bibinfo{year}{2015}\natexlab{}.
\newblock \showarticletitle{Session-based Recommendations with Recurrent Neural Networks}.
\newblock \bibinfo{journal}{\emph{arXiv preprint arXiv:1511.06939}} (\bibinfo{year}{2015}).
\newblock


\bibitem[Kang and McAuley(2018)]%
        {kang2018self}
\bibfield{author}{\bibinfo{person}{Wang-Cheng Kang} {and} \bibinfo{person}{Julian McAuley}.} \bibinfo{year}{2018}\natexlab{}.
\newblock \showarticletitle{Self-attentive sequential recommendation}. In \bibinfo{booktitle}{\emph{Proc. ICDM}}.
\newblock


\bibitem[Kipf and Welling(2016)]%
        {kipf2016semi}
\bibfield{author}{\bibinfo{person}{Thomas~N Kipf} {and} \bibinfo{person}{Max Welling}.} \bibinfo{year}{2016}\natexlab{}.
\newblock \showarticletitle{Semi-supervised classification with graph convolutional networks}.
\newblock \bibinfo{journal}{\emph{arXiv preprint arXiv:1609.02907}} (\bibinfo{year}{2016}).
\newblock


\bibitem[Park et~al\mbox{.}(2023)]%
        {park2023toward}
\bibfield{author}{\bibinfo{person}{Seongmin Park}, \bibinfo{person}{Mincheol Yoon}, \bibinfo{person}{Jae-woong Lee}, \bibinfo{person}{Hogun Park}, {and} \bibinfo{person}{Jongwuk Lee}.} \bibinfo{year}{2023}\natexlab{}.
\newblock \showarticletitle{Toward a Better Understanding of Loss Functions for Collaborative Filtering}. In \bibinfo{booktitle}{\emph{Proc. CIKM}}. \bibinfo{pages}{2034--2043}.
\newblock


\bibitem[Rendle et~al\mbox{.}(2009)]%
        {rendle2012bpr}
\bibfield{author}{\bibinfo{person}{Steffen Rendle}, \bibinfo{person}{Christoph Freudenthaler}, {et~al\mbox{.}}} \bibinfo{year}{2009}\natexlab{}.
\newblock \showarticletitle{BPR: Bayesian personalized ranking from implicit feedback}. In \bibinfo{booktitle}{\emph{Proc. UAI}}.
\newblock


\bibitem[Sun et~al\mbox{.}(2019)]%
        {sun2019bert4rec}
\bibfield{author}{\bibinfo{person}{Fei Sun}, \bibinfo{person}{Jun Liu}, {et~al\mbox{.}}} \bibinfo{year}{2019}\natexlab{}.
\newblock \showarticletitle{BERT4Rec: Sequential recommendation with bidirectional encoder representations from transformer}. In \bibinfo{booktitle}{\emph{Proc. CIKM}}.
\newblock


\bibitem[Veli{\v{c}}kovi{\'c} et~al\mbox{.}(2018)]%
        {velivckovic2017graph}
\bibfield{author}{\bibinfo{person}{Petar Veli{\v{c}}kovi{\'c}}, \bibinfo{person}{Guillem Cucurull}, \bibinfo{person}{Arantxa Casanova}, \bibinfo{person}{Adriana Romero}, \bibinfo{person}{Pietro Lio}, {and} \bibinfo{person}{Yoshua Bengio}.} \bibinfo{year}{2018}\natexlab{}.
\newblock \showarticletitle{Graph attention networks}. In \bibinfo{booktitle}{\emph{Proceedings of the ICLR}}.
\newblock


\bibitem[Wang et~al\mbox{.}(2024)]%
        {10.1145/3627673.3679535}
\bibfield{author}{\bibinfo{person}{Chen Wang}, \bibinfo{person}{Liangwei Yang}, \bibinfo{person}{Zhiwei Liu}, \bibinfo{person}{Xiaolong Liu}, \bibinfo{person}{Mingdai Yang}, \bibinfo{person}{Yueqing Liang}, {and} \bibinfo{person}{Philip~S. Yu}.} \bibinfo{year}{2024}\natexlab{}.
\newblock \showarticletitle{Collaborative Alignment for Recommendation} \emph{(\bibinfo{series}{CIKM '24})}. \bibinfo{publisher}{Association for Computing Machinery}, \bibinfo{address}{New York, NY, USA}.
\newblock
\showISBNx{9798400704369}


\bibitem[Wang et~al\mbox{.}(2022b)]%
        {wang2022towards}
\bibfield{author}{\bibinfo{person}{Chenyang Wang}, \bibinfo{person}{Yuanqing Yu}, {et~al\mbox{.}}} \bibinfo{year}{2022}\natexlab{b}.
\newblock \showarticletitle{Towards Representation Alignment and Uniformity in Collaborative Filtering}. In \bibinfo{booktitle}{\emph{Proc. SIGKDD}}.
\newblock


\bibitem[Wang and Isola(2020)]%
        {wang2020understanding}
\bibfield{author}{\bibinfo{person}{Tongzhou Wang} {and} \bibinfo{person}{Phillip Isola}.} \bibinfo{year}{2020}\natexlab{}.
\newblock \showarticletitle{Understanding contrastive representation learning through alignment and uniformity on the hypersphere}. In \bibinfo{booktitle}{\emph{Proc. ICML}}.
\newblock


\bibitem[Wang et~al\mbox{.}(2019)]%
        {wang2019neural}
\bibfield{author}{\bibinfo{person}{Xiang Wang}, \bibinfo{person}{Xiangnan He}, \bibinfo{person}{Meng Wang}, \bibinfo{person}{Fuli Feng}, {and} \bibinfo{person}{Tat-Seng Chua}.} \bibinfo{year}{2019}\natexlab{}.
\newblock \showarticletitle{Neural graph collaborative filtering}. In \bibinfo{booktitle}{\emph{Proceedings of the 42nd international ACM SIGIR conference on Research and development in Information Retrieval}}. \bibinfo{pages}{165--174}.
\newblock


\bibitem[Wang et~al\mbox{.}(2022a)]%
        {wang2022multi}
\bibfield{author}{\bibinfo{person}{Ziyang Wang}, \bibinfo{person}{Huoyu Liu}, \bibinfo{person}{Wei Wei}, \bibinfo{person}{Yue Hu}, \bibinfo{person}{Xian-Ling Mao}, \bibinfo{person}{Shaojian He}, \bibinfo{person}{Rui Fang}, {and} \bibinfo{person}{Dangyang Chen}.} \bibinfo{year}{2022}\natexlab{a}.
\newblock \showarticletitle{Multi-level contrastive learning framework for sequential recommendation}. In \bibinfo{booktitle}{\emph{Proceedings of the 31st ACM International Conference on Information \& Knowledge Management}}. \bibinfo{pages}{2098--2107}.
\newblock


\bibitem[Xu et~al\mbox{.}(2018)]%
        {xu2018powerful}
\bibfield{author}{\bibinfo{person}{Keyulu Xu}, \bibinfo{person}{Weihua Hu}, \bibinfo{person}{Jure Leskovec}, {and} \bibinfo{person}{Stefanie Jegelka}.} \bibinfo{year}{2018}\natexlab{}.
\newblock \showarticletitle{How powerful are graph neural networks?}
\newblock \bibinfo{journal}{\emph{arXiv preprint arXiv:1810.00826}} (\bibinfo{year}{2018}).
\newblock


\bibitem[Yang et~al\mbox{.}(2023)]%
        {yang2023graph}
\bibfield{author}{\bibinfo{person}{Liangwei Yang}, \bibinfo{person}{Zhiwei Liu}, \bibinfo{person}{Chen Wang}, \bibinfo{person}{Mingdai Yang}, \bibinfo{person}{Xiaolong Liu}, \bibinfo{person}{Jing Ma}, {and} \bibinfo{person}{Philip~S Yu}.} \bibinfo{year}{2023}\natexlab{}.
\newblock \showarticletitle{Graph-based alignment and uniformity for recommendation}. In \bibinfo{booktitle}{\emph{Proceedings of the 32nd ACM International Conference on Information and Knowledge Management}}. \bibinfo{pages}{4395--4399}.
\newblock


\bibitem[Yang et~al\mbox{.}(2024)]%
        {DBLP:conf/wsdm/YangLYLWPY24}
\bibfield{author}{\bibinfo{person}{Mingdai Yang}, \bibinfo{person}{Zhiwei Liu}, \bibinfo{person}{Liangwei Yang}, \bibinfo{person}{Xiaolong Liu}, \bibinfo{person}{Chen Wang}, \bibinfo{person}{Hao Peng}, {and} \bibinfo{person}{Philip~S. Yu}.} \bibinfo{year}{2024}\natexlab{}.
\newblock \showarticletitle{Unified Pretraining for Recommendation via Task Hypergraphs}. In \bibinfo{booktitle}{\emph{Proceedings of the 17th {ACM} International Conference on Web Search and Data Mining, {WSDM} 2024, Merida, Mexico, March 4-8, 2024}}, \bibfield{editor}{\bibinfo{person}{Luz~Angelica Caudillo{-}Mata}, \bibinfo{person}{Silvio Lattanzi}, \bibinfo{person}{Andr{\'{e}}s~Mu{\~{n}}oz Medina}, \bibinfo{person}{Leman Akoglu}, \bibinfo{person}{Aristides Gionis}, {and} \bibinfo{person}{Sergei Vassilvitskii}} (Eds.). \bibinfo{publisher}{{ACM}}, \bibinfo{pages}{891--900}.
\newblock
\urldef\tempurl%
\url{https://doi.org/10.1145/3616855.3635811}
\showDOI{\tempurl}


\bibitem[Zhao et~al\mbox{.}(2021)]%
        {zhao2021recbole}
\bibfield{author}{\bibinfo{person}{Wayne~Xin Zhao}, \bibinfo{person}{Shanlei Mu}, \bibinfo{person}{Yupeng Hou}, \bibinfo{person}{Zihan Lin}, \bibinfo{person}{Yushuo Chen}, \bibinfo{person}{Xingyu Pan}, \bibinfo{person}{Kaiyuan Li}, \bibinfo{person}{Yujie Lu}, \bibinfo{person}{Hui Wang}, \bibinfo{person}{Changxin Tian}, {et~al\mbox{.}}} \bibinfo{year}{2021}\natexlab{}.
\newblock \showarticletitle{Recbole: Towards a unified, comprehensive and efficient framework for recommendation algorithms}. In \bibinfo{booktitle}{\emph{proceedings of the 30th acm international conference on information \& knowledge management}}. \bibinfo{pages}{4653--4664}.
\newblock


\end{thebibliography}



\end{document}